\documentclass[hyper]{JHEP3}
\usepackage{latexsym,amssymb,amsmath,epsfig,epic,eepic}
\newcommand{\beq}{\begin{equation}} 
\newcommand{\eeq}{\end{equation}}
\newcommand{\beqs}{\begin{eqnarray}}
\newcommand{\eeqs}{\end{eqnarray}}

\newcommand{\nn}{\nonumber}

\newcommand{\ov}{\overline}

\newcommand{\ph}{\phantom}

\title{Superpotentials From Stringy Instantons \\
Without Orientifolds}

\author{ Christoffer Petersson\\
Fundamental Physics\\ Chalmers
University of Technology \\ SE 412 96 G\"oteborg, Sweden\\
\vspace{0.1cm}

\email{chrpet@chalmers.se}\\}

\abstract{In this paper we show that it is possible to derive non-perturbative superpotential terms from a stringy instanton without introducing orientifold planes. The instanton is realized by a Euclidean D brane wrapping a non-trivial cycle upon which we also wrap a single space-filling D brane. The standard problem of unwanted neutral fermionic zero modes is evaded by the appearance of couplings to charged bosonic zero modes in the instanton moduli action. Since the Euclidean D brane wraps a cycle which is not associated to any low energy gauge dynamics, it can not be interpreted as an ordinary gauge instanton, but rather as a stringy one. By considering such a brane configuration at an orbifold singularity, we can explicitly evaluate the instanton moduli space integral and find a holomorphic superpotential term with the structure of a baryonic mass term.}

\begin{document}
\setcounter{section}{0}

\renewcommand{\thefootnote}{\arabic{footnote}}
\setcounter{footnote}{0} \setcounter{page}{1}

\tableofcontents

\section{Introduction}

There has been interesting recent developments in the context of string theory realizations of instanton effects in gauge theories \cite{Witten:1995gx,Douglas:1995bn,Witten:1996bn,Ganor:1996pe,Green:2000ke,Billo:2002hm}. Non-perturbative superpotential terms which are known to be generated by a single instanton, such as the ADS superpotential in $\mathcal{N}$=1 SQCD for the case $N_f=N_c -1$, have been explicitly derived using boundary conformal field theory \cite{Akerblom:2006hx,Dorey:2002ik,Beasley:2004ys}. Such a gauge instanton can be realized by a Euclidean D brane (ED brane) wrapped on a non-trivial cycle upon which the space-filling D branes that make up the gauge group, in an engineered SQCD theory, have also been wrapped. 

It has further been shown that certain stringy realizations lead to non-perturbative superpotential terms generated by instantons which do not admit an obvious interpretation from an ordinary gauge theory point of view \cite{Blumenhagen:2006xt,Ibanez:2006da,Florea:2006si,Abel:2006yk,Bianchi:2007fx,Cvetic:2007ku,Argurio:2007vqa,Bianchi:2007wy,Ibanez:2007rs,Blumenhagen:2007sm}. Such cases are realized by an ED brane wrapping a cycle which has no gauge dynamics associated to it and is called a stringy instanton. It has been shown that it is possible to generate phenomenologically interesting superpotential terms, such as Majorana mass terms for right handed neutrinos \cite{Blumenhagen:2006xt,Ibanez:2006da,Cvetic:2007ku,Antusch:2007jd}.

In the stringy case of an ED brane on a cycle which has no space-filling D brane already wrapped on it, there generically arises a problem due to an excess of neutral fermionic zero modes. The reason is that for an open string, with both endpoints on an ED brane that wraps an otherwise ``unoccupied'' cycle, does not feel the presence of the space-filling D branes and therefore, gives rise to 4 fermionic massless modes, corresponding to the supertranslations broken by the ED brane in an $\mathcal{N}$=2 background. This implies that instead of the 2 (Goldstino) zero modes required for the generation of a superpotential term we get additional fermionic moduli fields that do not appear in the moduli action and hence make the instanton moduli space integral vanish. The standard way to get rid of these unwanted fermionic zero modes is to introduce an orientifold plane and thereby project these extra modes out \cite{Argurio:2007qk,Argurio:2007vqa,Bianchi:2007wy,Franco:2007ii,Ibanez:2007tu}. There have also been investigations concerning the possibility of lifting these fermionic moduli fields by including background fluxes together with gauge flux on the world volume of the ED brane \cite{Blumenhagen:2007bn,Martucci:2005rb,Bergshoeff:2005yp,Haack:2006cy}. 

In this paper, we will show that it is in fact possible to generate non-perturbative superpotential terms from stringy instantons without introducing orientifolds or taking closed string modes into account. As our main focus will be to show how the problem of unwanted neutral fermionic zero modes can be evaded we will throughout the paper only be considering local configurations and not be concerned with global issues such as cancellation of D brane induced tadpoles which in general require the presence of orientifold planes.  We will work in a type IIB $\mathbb{Z}_2 \times \mathbb{Z}_2$ orbifold background, where we can use a simple CFT description when studying the interactions between the massless modes of the open strings stretching between the various branes. Although the $\mathcal{N}$=1 non-chiral world volume gauge theory this orbifold background gives rise to is not of particular phenomenological importance, we believe that the results we obtain are quite general and applicable to many other D brane systems in various Calabi-Yau backgrounds \cite{Franco:2005zu,Bertolini:2005di,Berenstein:2005xa}.

The key point in generating the non-perturbative superpotential term will be to consider branes wrapping 3 different 2-cycles. We wrap $N_1$ D5 branes on the first cycle, $N_2$ on the second and a single D5 brane on the third, $N_3 =1$. By also wrapping an ED1 on the third cycle we are in the situation where we have an instanton which is not associated to any low energy gauge dynamics, since there is only an IR free U(1) factor here. However, due to the presence of bosonic zero modes between the ED1 brane and the single D5 brane, there will appear couplings in the effective instanton moduli action which involve the unwanted fermionic zero modes. The integration over these extra fermionic zero modes imposes constraints on the remaining moduli fields, analogous to the  fermionic ADHM \cite{Atiyah:1978ri} constraints one imposes on the moduli fields of an conventional gauge instanton, and we are left with an integral which has the correct number (two) of neutral fermionic zero modes to make up the integration measure over chiral superspace.  When performing the remaining integrations, we find that a holomorphic superpotential term with the structure of a baryonic mass term is obtained for the case when $N_1=N_2$, without introducing any orientifolds. We regard the computation done in this paper as an explicit confirmation of the expectations raised in \cite{Aganagic:2007py,GarciaEtxebarria:2007zv}
for related configurations. We will see that, by including the baryonic mass term as a non-perturbative part of the superpotential, the R-charge assignment of the chiral superfields is uniquely fixed in the non-chiral quiver gauge theory under consideration and moreover, an axial U(1) symmetry is broken. Note however that, in other configurations where orientifolds were used, it has been shown that the inclusion of such a baryonic mass term, in some instances, leads to dynamical supersymmetry breaking \cite{Argurio:2007qk,Aharony:2007db,Buican:2007is,Aharony:2007pr}.  
 
The plan of this paper is as follows. In section \ref{orbi} we first review the field content of the $k=1$ instanton sector of the $\mathcal{N}$=4 SYM theory, realized by a D(-1) instanton in the world volume of $N$ D3 branes, and then perform a $\mathbb{Z}_2 \times \mathbb{Z}_2$ orbifold projection to obtain our $\mathcal{N}$=1 SQCD-like gauge theory with one instanton. We review the open string spectrum for such a configuration and write down the corresponding effective instanton moduli action in the ADHM limit. In section \ref{pre} we discuss the prefactor of the possibly generated non-perturbative superpotential term and also the power to which the chiral superfields should appear in such a term. In section \ref{eva} we explicitly evaluate the moduli space integral for the configuration when one of the cycles is wrapped by a single space-filling D brane together with an instanton ED brane, and we find a non-vanishing holomorphic result. In section \ref{imp} we give a brief discussion of the implications on the gauge theory dynamics we should expect from the non-perturbative superpotential term found in section \ref{eva}.

\section{The Orbifold Projection of the $k=1$ Instanton Sector of $\mathcal{N}$=4 SYM}
\label{orbi}

In this section we will review the open string spectrum for a system with $N$ D3 branes and one D(-1) brane ($k=1$) in a type IIB background \cite{Dorey:2002ik,Green:2000ke,Billo:2002hm,Bianchi:2007ft}. Since we are interested in instanton calculus we Wick rotate our ten dimensional Minkowski spacetime, according to \cite{Dorey:2002ik}. 

In the gauge sector the massless modes of the open strings, with its endpoints attached to two of the $N$ D3 branes, form an $\mathcal{N}$=4 SYM multiplet \cite{Witten:1995im}. In the NS sector, we obtain the gauge field $A^{\mu}$ from the oscillators with spacetime indices pointing along the D3 brane. The oscillators pointing in the 3 complex directions transverse to the D3 brane give, in $\mathcal{N}$=1 language, the three chiral superfields $\Phi^{1,2,3}$. All fields in the gauge sector are in the adjoint representation of U($N$).

The fields in the neutral sector correspond to the zero modes of the strings with both ends on the D(-1) brane. These fields do not transform under the gauge group of the D3 branes but instead in the adjoint representation of the instanton gauge group which, for a single ($k=1$) instanton, is simply U(1). In the same way as the $\mathcal{N}$=4 SYM theory in 4 dimensions can be obtained from a dimensional reduction of the $\mathcal{N}$=1 SYM theory in ten dimensions \cite{Brink:1976bc}, the neutral sector can be obtained by continuing the reduction down to zero dimensions. We denote the four bosonic moduli fields, longitudinal to the D3 branes world volume, by $a^\mu$. Also, from the oscillators with spacetime indices in the directions transverse to the D3 branes we get another six bosonic moduli fields which we will however not be concerned with since they will be projected out by the orbifold projection in the configurations we will consider later on.  In the R sector, the fermionic zero-modes are denoted $M^{\alpha A}$ and
$\lambda_{\dot\alpha A}$, where $\alpha$/$\dot\alpha$ denote SO(4) Weyl spinor indices of positive/negative chirality transforming in the fundamental representation under the respective factor of SU(2)$\times$SU(2)$\cong$SO(4) and $A$ upstairs/downstairs denote SO(6) Weyl spinor indices of negative/positive chirality which transform in the fundamental/anti-fundamental representation of SU(4)$\cong$SO(6). Hence, the presence of the D3 branes have broken the Euclidean Lorentz group SO(10) to SO(4)$\times$SO(6) and the ten dimensional chirality of
both fermionic fields have been chosen to be negative. We will also introduce a triplet of auxiliary fields $D^c$  that can be used to decouple quartic moduli interactions and linearize supersymmetry transformations but most importantly, it is crucial in order to recover the standard ADHM results in the field theory ($\alpha'\to 0$) limit \cite{Dorey:2002ik,Billo:2002hm}. Since we will only be dealing with a single instanton, the neutral sector fields are not matrix valued.

 The charged sector fields come from the zero-modes of the strings
stretching between the D(-1) brane and one of the $N$ D3 branes.  For each such open string we have two conjugate sectors distinguished by the orientation of the string.
In the NS sector, where the world-sheet fermions have opposite moding
compared to the bosons, we obtain a bosonic SO(4) Weyl spinor $\omega_{\dot\alpha}$ in the
first four directions where the GSO projection picks out the negative
chirality. In the conjugate sector, we will get an independent bosonic SO(4) Weyl 
spinor $\bar\omega_{\dot\alpha}$ of the same chirality.  In
the R sector, we obtain two independent SO(6) Weyl spinors $\mu^A$ and $\bar\mu^A$, one for each conjugate sector, with chirality fixed by the GSO projection such that both spinors transform in the fundamental representation of SU(4)$\cong$SO(6). Note that the moduli fields in the charged sector with(out) a ``bar'' are in the anti-fundamental (fundamental) representation of U($N$), the world volume gauge group of the D3 branes.

Let us now perform an orbifold projection \cite{Douglas:1996sw,Morrison:1998cs} on the configuration described above. We will choose  the orbifold group to be $\mathbb{Z}_2 \times \mathbb{Z}_2$, which will give us a non-chiral $\mathcal{N}$=1 quiver gauge theory \cite{Bertolini:2001gg}. Since the orbifold projection was done in detail in \cite{Argurio:2007vqa} we will here only state the action of the two generators $h_1$ and $h_2$ of the two $\mathbb{Z}_2$ (see Table 1) 
\begin{table} 
\begin{center}
\begin{tabular}{c||c|c|c|}
&$z^1$ & $z^2$ & $z^3$ \\ \hline   \hline $h_1$  &$z^1$ & $-z^2$ & $-z^3$ \\  \hline $h_2$ &$-z^1$ &
$z^2$ & $-z^3$ \\  \hline 
\end{tabular}
\caption{\small The action of the orbifold generators.}
\end{center}
\end{table}
and its regular representations $\gamma(h)$ acting on the Chan-Paton factors,
\beq
\gamma(h_1)  =\begin{pmatrix}1 & 0 & 0   & 0   \cr  0 & 1 & 0   & 0 \cr 0 & 0
& -1   &  0 \cr  0   & 0   & 0 & -1 \cr\end{pmatrix}~~~,~~~
\gamma(h_2) =\begin{pmatrix}1 & 0 & 0   & 0   \cr   0 & -1 & 0   & 0 \cr 0 & 0
& 1   &  0 \cr  0   & 0   & 0 & -1 \cr\end{pmatrix}~
\label{chanpatonz2z2}
\eeq
where the 1's denote $N_\ell \times N_\ell$ unit matrices, $\ell=1,...,4$ and $\sum_{\ell =1}^{4}N_\ell =N$. For a review on fractional branes, see \cite{Bertolini:2003iv}.

\subsection{The Gauge Sector}

In the gauge sector, the orbifold projection implies that the vector superfields are block diagonal matrices of different size $(N_1,N_2,N_3,N_4)$, one for
each node of the quiver. Since we will throughout the paper never occupy node 4 with fractional D3 branes, we set $N_4 =0$ from now on. Thus, our gauge group is U($N_{1}$)$\times$U($N_{2}$)$\times$U($N_{3}$).
The three chiral superfields $\Phi^i$ will have the following form
\begin{equation}
\Phi^1 = \begin{pmatrix}0 & \Phi_{12} &
0   & 0   \cr  \Phi_{21} & 0 & 0   & 0 \cr 0 & 0   & 0   &  \mathbf{0} \cr  0
& 0   & \mathbf{0} & 0 \cr\end{pmatrix}, \Phi^2 = \begin{pmatrix}0 & 0
& \Phi_{13} &  0  \cr     0   & 0   & 0   & \mathbf{0} \cr \Phi_{31} & 0   & 0
& 0  \cr  0   & \mathbf{0} & 0 & 0 \cr\end{pmatrix},  \Phi^3 =
\begin{pmatrix}0 & 0   & 0   & \mathbf{0} \cr  0   & 0 & \Phi_{23} & 0   \cr
0 & \Phi_{32} & 0   & 0   \cr  \mathbf{0} & 0   & 0   & 0 \cr\end{pmatrix} 
\label{structure}
\end{equation}
where the non-zero entries $\Phi_{\ell m}$ denote chiral superfields transforming in the fundamental representation of gauge group U($N_{\ell}$) and in the anti-fundamental of gauge group U($N_m$). The associated quiver diagram is displayed in Figure 1.

\begin{figure}
\begin{center}
\includegraphics[width=75mm]{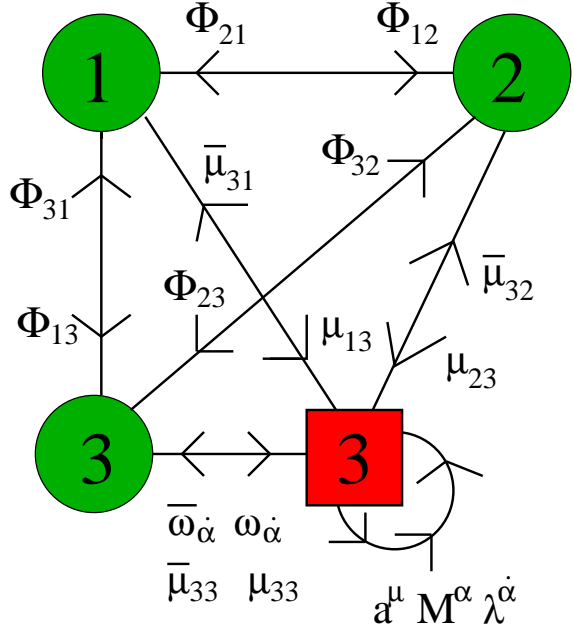}
\caption{\small{The $\mathbb{Z}_2 \times \mathbb{Z}_2$ orbifold quiver gauge theory where the fractional D3 branes (green circles) have been given rank assignment ($N_1$,$N_2$,$N_3$,0). We have also included all neutral and charged zero modes of the fractional instanton (red square) which is located at node 3, together with $N_3$ fractional D3 branes.}}
\label{gauge}
\end{center}
\end{figure}

\subsection{The Neutral Sector}

The Chan-Paton structure for the fractional D(-1) instantons will be the same as for the gauge sector. However, since we will only be considering a single instanton at node 3, all off diagonal
neutral modes are absent, as they connect instantons at two
distinct nodes. Thus, we keep only the third diagonal component in the 4$\times$4 Chan-Paton matrix of the neutral fields, corresponding to the case where $k_3=1$ and $k_1=k_2=k_4=0$. The 4 bosonic zero modes in the NS sector that remains, corresponding to the location of this fractional instanton in the world volume of the fractional D3 branes, will (also here) be denoted by $a^\mu$.  

 For the fermionic moduli fields $M^{\alpha A}$ and $\lambda_{\dot\alpha A}$ we can choose a representation of the Dirac matrices such that
$M^{\alpha A}$ and $\lambda_{\dot\alpha A}$ for $A=1,2,3$ have
the Chan-Paton structure of (\ref{structure}), while for $A=4$ they are block diagonal \cite{Argurio:2007vqa}. Thus, since we are only interested in a diagonal component of the Chan-Paton matrix, the only neutral fields that survive the projection are those with SU(4) index 4. We denote the third component of these remaining fermionic moduli fields by $M^{\alpha}$ and $\lambda^{\dot\alpha}$.

\subsection{The Charged Sector}

 The charged sector is now described by the open strings stretching
from the fractional instanton at node 3 to the fractional D3 branes, and vice versa. Since the charged bosonic moduli fields do not carry indices which point in any of the directions the orbifold acts on, the Chan-Paton factor will have a block diagonal structure and thus we will only find surviving fields among the zero modes of the open strings between the fractional D(-1)${}_3$ and the D3 branes at node 3.\footnote{The block diagonal structure of the Chan-Paton factors can also be understood from the fact that the charged open strings stretching from the fractional D(-1$)_{3}$ instanton to one of the fractional D3 branes, which is not at node 3, would behave as 8 Dirichlet-Neumann strings with massive NS ground state since the ED1 and D5 then wrap different 2-cycles.} Hence, we obtain 4$N_3$ bosonic zero modes $\omega_{\dot\alpha}, ~\bar\omega_{\dot\alpha}$. 

The charged fermionic zero modes $\mu^A$ and $\bar\mu^A$  will display the same
structure as in (\ref{structure}) for $A=1,2,3$, but they will be block
diagonal for $A=4$. This means that between the fractional instanton and the $N_3$ D3 branes at node 3 we have $2N_3$ fermionic zero-modes $\mu_{33}$ and $\ov\mu_{33}$. As the SU(4) indices will not be written explicitly, we simply note that these charged fermions correspond to the SU(4) index 4. Between the instanton and the $N_1$ D3 branes at node 1 there are $2N_1$ fermionic zero-modes $\mu_{13}$ and $\bar\mu_{31 }$, corresponding to SU(4) index 2.  Finally, between the instanton and the $N_2$ D3 branes at node 2, we have $2N_2$ fermionic zero-modes $\mu_{23}$ and $\bar\mu_{32 }$, with SU(4) index 3.  Note that, in order to ease the notation, we do not write out the fundamental indices of the charged moduli fields (without  a ``bar''), corresponding to the gauge group of the fractional D3 node the string stretches from, and similarly for the anti-fundamental indices of the charged fields (with a ``bar'') stretching to the D3 branes.

\subsection{The Moduli Space Integral}

We can now calculate tree level open string scattering amplitudes by inserting the vertex operators for the moduli fields at the boundary of a disk, corresponding to the world volume of the open string. 
In order to recover the standard ADHM result for an ordinary gauge instanton we take the ``ADHM limit'', implying that we, in addition to taking the field theory limit $\alpha' \to 0$, perform a particular rescaling of the moduli fields, see \cite{Billo:2002hm}, and then send $g_0 \to \infty$ while keeping the 4-dimensional D3 brane world volume gauge coupling fixed. By summing over all amplitudes that survive this limit we recover the following effective instanton moduli action for a single fractional instanton, 
\begin{equation}
\label{S1}
S_1 = i \left(\ov\mu_{33} \omega_{ \dot\alpha} +
\ov\omega_{\dot\alpha} \mu_{33}  
 \right)\! \lambda^{\dot\alpha }
- i D^c \big( \ov\omega^{ \dot\alpha}
(\tau^c)^{\dot\beta}_{\dot\alpha} \omega_{\dot\beta} \big) ~
\end{equation}
and the interaction terms between the charged sector and the chiral superfields are given by 
\begin{eqnarray}
\label{S21}
S_2 &=& \frac{1}{2} \overline{\omega}^{\dot{\alpha}} \big( \Phi_{31} \overline\Phi_{13} +\overline\Phi_{31} \Phi_{13} +\Phi_{32} \overline\Phi_{23} +\overline\Phi_{32} \Phi_{23}  \big)  \omega_{\dot{\alpha}} \nn \\
&&  +\frac{i}{2}\overline{\mu}_{31}\ov\Phi_{13}\mu_{33} -\frac{i}{2}\overline{\mu}_{33}\ov\Phi_{31}\mu_{13}+\frac{i}{2}\overline{\mu}_{32}\ov\Phi_{23}\mu_{33}-\frac{i}{2}\overline{\mu}_{33}\ov\Phi_{32}\mu_{23} \nn \\
&&-\frac{i}{2}\overline{\mu}_{32}\Phi_{21}\mu_{13} + \frac{i}{2}\overline{\mu}_{31}\Phi_{12}\mu_{23} ~
\end{eqnarray}
where we have both holomorphic and anti-holomorphic couplings. All terms in (\ref{S1}) and (\ref{S21}) can be obtained by performing the $\mathbb{Z}_2 \times \mathbb{Z}_2$ orbifold projection of the parent $k=1$ instanton sector of the $\mathcal{N}=4$ theory \cite{Argurio:2007vqa}. 

We will throughout the paper assume that it makes sense to take the ADHM limit of the instanton moduli action. For a conventional gauge instanton, this is the limit that yields, first of all, the standard ADHM measure on the instanton moduli space of the $\mathcal{N}=4$ D3 world volume gauge theory before the orbifold projection \cite{Dorey:2002ik,Billo:2002hm}, but also the one instanton generated ADS superpotential of the $\mathcal{N}=1$ fractional D3 world volume gauge theory after the $\mathbb{Z}_2 \times \mathbb{Z}_2$ orbifold projection \cite{Argurio:2007vqa}. Even though we will later on be concerned with instantons that do not admit an obvious interpretation in terms of ordinary commutative gauge theory they will however have similarities with ordinary gauge instantons since they, for example, have charged bosonic moduli associated to them.  

As suggested by \cite{Blumenhagen:2006xt,Dorey:2002ik}, if a non-perturbative superpotential is generated in the configuration described above, its form can be obtained by evaluating the moduli space integral 
\begin{equation}
\label{SW}
S_W = \mathcal{C} \int d\{ a,M,\lambda, D, \omega,\ov\omega, \mu,\ov\mu  \}e^{-S_1 -S_2}~~.
\end{equation} 
We will in the following two sections, first discuss the prefactor $\mathcal{C}$, here inserted in order to compensate for the dimension of the moduli space measure, and then explicitly evaluate the integral (\ref{SW}).

\section{Determining the Prefactor}
\label{pre}

In this section we will start by considering the case when there are $N_3 >1$ fractional D3 branes together with the fractional instanton at node 3, corresponding to an ordinary gauge instanton associated to the gauge group at node 3. We then turn to the stringy instanton case, $N_3=1$, and discuss the structure we expect the generated superpotential term to have, using dimensional analysis.   

\subsection{The Gauge Instanton}

In order to check that the action term in (\ref{SW}) is dimensionless we need to know the scaling dimension of all the moduli fields that appear in the instanton measure. This can be obtained by demanding a dimensionless moduli action in (\ref{S1}) \cite{Billo:2002hm,Bianchi:2007wy},
\begin{eqnarray}
[a^\mu ] & = & [\omega_{\dot{\alpha}} ]=[\overline{\omega}_{\dot{\alpha}} ]=M^{-1}_{s} ~,~ [D^c ]=M^{2}_{s} \nn \\
\big[ M^{\alpha} \big] & = & [\mu ]=[\overline{\mu}]= M^{-1/2}_{s}  ~,~ [\lambda^{ \dot\alpha } ]  =  M^{3/2}_{s}  ~.
\end{eqnarray}
where $M_s =1/\sqrt{\alpha'}$. For the configuration under consideration, with fractional D3 brane rank assignment ($N_1$,$N_2$,$N_3$,0) and fractional D(-1) rank assignment (0,0,1,0), the dimension of the instanton measure is given by
\begin{eqnarray}
\label{dimmeas}
\Big[  d\{ a,M,\lambda, D, \omega,\ov\omega, \mu,\ov\mu  \} \Big] &=& M_{s}^{-(n_{a } -\frac{1}{2}n_{M} +\frac{3}{2}n_{\lambda} -2 n_{D}+ n_{\omega,\ov\omega} - \frac{1}{2}n_{\mu , \ov\mu } )} \nn \\
&=& M_{s}^{-( n_{\omega,\ov\omega} - \frac{1}{2}n_{\mu , \ov\mu } )} = M_{s}^{-(3N_3 -N_1 -N_2)}=M_{s}^{-\beta_3}
\end{eqnarray}
since we have $n_{\omega,\ov\omega}=4N_3$ charged bosons ($\omega_{\dot\alpha}$ and $\ov\omega_{\dot\alpha}$) and  $n_{\mu , \ov\mu }=2N_3 +2N_1 +2N_2$ charged fermions ($\mu_{33}$, $\ov\mu_{33}$, $\mu_{13}$, $\ov\mu_{31}$, $\mu_{23}$ and $\ov\mu_{32}$) in addition to the instanton gauge field $a^\mu$ ($n_{a }=4$), its superpartners $M^{\alpha}$ and $\lambda^{ \dot\alpha }$ ($n_{M}=2$ and $n_{\lambda}=2$) and the $n_D =3$ auxiliary fields $D^c$. In (\ref{dimmeas}) we have denoted the dimension of the instanton measure by $\beta_3$ since we recognize it as the one loop $\beta$-function coefficient for the gauge coupling constant $g_3$ of the $\mathcal{N}$=1 U($N_3$) vector multiplet, together with the contribution from $N_1 +N_2$ generations of bi-fundamental chiral superfields, $\beta_3 =3N_3 -N_1 -N_2$ \cite{Taylor:1982bp,Affleck:1983mk}. The one loop $\beta$-function coefficient $\beta_3$ can also be obtained by calculating the annulus vacuum amplitude 
for the open strings between the D$(\mathrm{-}1)_3$ instanton and the fractional D3 branes \cite{Blumenhagen:2006xt,Akerblom:2006hx,Bianchi:2007wy,Akerblom:2007uc,Billo:2007py,Billo:2007sw}. This one loop running of the gauge coupling constant $g_3$ is due to the massless states circulating the loop and because we take the strictly local point of view, there are no threshold corrections  due to higher string states \cite{Lust:2003ky,Akerblom:2007np,Camara:2007dy}. The absence of higher string state contribution together with the fact that we are performing the integration over the instanton zero modes explicitly in (\ref{SW}) implies that, in order to not overcount, there is no contribution from the annulus diagrams to the prefactor $\mathcal{C}$ in (\ref{SW}). 


The only missing piece of the prefactor is obtained by taking into account the vacuum disk diagrams which have their boundaries completely on the fractional instanton at node 3 and contribute by an exponential of the topological normalization of a D$(\mathrm{-}1)_3$ disk\footnote{Note that such a vacuum disk amplitude is also given by minus the classical instanton action \cite{Polchinski:1994fq}.}, $-\frac{8\pi^2}{g_{3}^{2}}$,
where again $g_{3}$ is the U($N_3$) gauge coupling constant, at the string scale $M_s$. Thus, by combining the dimensionful factor $M_{s}^{\beta_3}$, compensating the dimension of the instanton moduli measure (\ref{dimmeas}), together with the vacuum disk exponential we can now identify the prefactor in (\ref{SW}) with the one loop renormalization group invariant scale $\Lambda$ of the U($N_3$) gauge theory on the world volume of the fractional D3 branes at node 3, 
\begin{equation}
\label{pref}
\mathcal{C}  = M_{s}^{\beta_3}~e^{-\frac{8\pi^2}{g_{3}^{2} }}=\Lambda^{\beta_3} ~
\end{equation}
where we have suppressed the $\theta$-angle dependence.


Since we can identify the neutral zero modes $a^\mu$ and $M^{\alpha}$ as coming from the super-translations broken by the fractional instanton we will henceforth denote them by $x^\mu =a^\mu$ and $\theta^{\alpha} =M^{\alpha}$. Thus, we can pull out these modes from the moduli integral in (\ref{SW}) and obtain the measure over chiral superspace $d^4 x d^2 \theta$. This allows us to determine to which power the chiral superfields will appear in the instanton generated superpotential term, 
\begin{equation}
\label{ }
S_W = \int  d^4 x d^2 \theta ~W_{\mathrm{np}}~
\end{equation}
where the non-perturbative superpotential is given by
\begin{equation}
\label{SW1}
W_{\mathrm{np}} = \Lambda^{\beta_3}  \int ~d\{ \lambda, D, \omega,\ov\omega, \mu,\ov\mu \}e^{-S_1 -S_2} \sim \Lambda^{\beta_3} \Phi^{-\beta_3 +3 } ~.
\end{equation} 
This is the usual form of the ($N_f =N_c -1$) ADS superpotential \cite{Taylor:1982bp,Affleck:1983mk}, which is generated by an instanton when $N_1 +N_2 =N_3 -1$, where $N_f =N_1+N_2$ is the number of flavors and $N_c =N_3$ is the number of colors. Using this constraint, we note that the power to which the chiral superfields in (\ref{SW1}) appear is negative, implying that the majority of fields will appear in the denominator, as  expected, since such a term is generated by a gauge instanton. In the remainder of this section, we will consider the $N_3 =1$ case, where the non-perturbative superpotential term is generated by an instanton which does no longer have an obvious gauge theory interpretation. 
 
 \subsection{The Stringy Instanton}

Let us now turn to the main focus of our study, which is the case when we only have a single fractional D3 brane at node 3, $N_3 =1$, where the instanton is located. There is no longer any low energy gauge dynamics associated with the third node since the U(1) factor is IR free. 

From the dimensional counting of the moduli measure in (\ref{dimmeas}) we see that, for $N_3 =1$, the coefficient to which dynamical scale $\Lambda$ in (\ref{pref}) appears is $(3-N_1 -N_2 )$. 
Although this coefficient can no longer be interpreted as an ordinary one loop $\beta$-function coefficient, we can conclude that if it was possible to generate a holomorphic superpotential term for this configuration, it would have the following structure,
\begin{equation}
\label{SW2}
W_{\mathrm{np}}^{\mathrm{s}} \sim 
 ~\Lambda_{\mathrm{string}}^{3-N_1 -N_2}~ \Phi^{N_1 +N_2}~.
 \end{equation} 
We have here labeled the scale $\Lambda$ with the subscript ``string'' in order to indicate the fact that it no longer has an ordinary gauge theory interpretation, but is of stringy origin. Since the power to which the chiral superfields in (\ref{SW2}) appear is positive we conclude that majority of fields will appear in the numerator and hence such a superpotential term can not be generated by an ordinary gauge instanton. Note that the only way to satisfy the ADS constraint $N_1+N_2=N_3-1$ for $N_3=1$ is to set the number of flavors $N_1=N_2=0$. Note also that the instanton disk vacuum amplitude can no longer be interpreted as the instanton classical action for an ordinary gauge instanton, but rather as the normalization of a D$(\mathrm{-}1)_3$ disk, since there exists no instanton solutions for ordinary commutative U(1) gauge theory. 

It is interesting to note that the coefficient $(3-N_1 -N_2 ) $ and the instanton disk vacuum amplitude obtained in this case have the same appearance as one would expect the one-loop $\beta$-function coefficient and the instanton action to have for a noncommutative U(1) gauge theory with $N_1+N_2$ flavors \cite{Martin:1999aq,Hayakawa:1999zf,Khoze:2000sy}. Moreover, it is known that noncommutativity in U($N$) gauge theories have a particular dramatic effect for the case $N=1$  since it is only on a noncommutative background that abelian gauge theories become non-trivial and allow for instanton solutions \cite{Nekrasov:1998ss}. Therefore, one might expect that the case under consideration is related to such configurations.\footnote{The author would like to thank Jose Francisco Morales for pointing this out.} 
We leave these issues for future work.

If we did not have any fractional D3 branes at all at node 3, only the chiral superfields 
$\Phi_{12}$ and $\Phi_{21}$ would exist \cite{Argurio:2007vqa}. In that case, there are no charged bosonic zero modes, the instanton moduli action in (\ref{S1}) vanishes and we only have the couplings in the last line of (\ref{S21}) left. Hence, if it was not for the two neutral fermionic $\lambda^{\dot\alpha}$-fields, the moduli space integral would yield a contribution of the form $\det [\Phi_{21}] \det [\Phi_{12}]$ for $N_1=N_2$, since the charged fermionic zero modes appear symmetrically. We note that this term  has the same dimension as the chiral superfields should have according to (\ref{SW2}). The problem with the case when there are no fractional D3 branes at node 3 is of course that the $\lambda^{\dot\alpha}$-modes make the integral vanish since they do not appear in the effective moduli action. 

As will be shown in the following section, a non-perturbative superpotential term like $\det [\Phi_{21}] \det [\Phi_{12}]$ is exactly what we find when we evaluate the moduli space integral for the case when we do have a single fractional D3 brane at node 3. In this case, there is no longer any problems with the 
unwanted $\lambda^{\dot\alpha}$-modes, since we now have charged bosonic zero modes which make these neutral fermions appear as Lagrange multipliers, implementing constraints completely analogous to the fermionic ADHM constraints one obtains for ordinary gauge instantons in the ADHM limit. The difference in this configuration is that the instanton here can not be interpreted as an ordinary gauge instanton, but rather as a stringy one.

\section{Evaluating the Moduli Space Integral}
\label{eva}

In this section, we will explicitly evaluate the instanton moduli space integral for the stringy $N_3=1$ configuration described above,
\begin{eqnarray}
\label{int}
W_{\mathrm{np}}^{\mathrm{s}}&=& \Lambda_{\mathrm{string}}^{3-N_1 -N_2} \int  d^3 D^c  d^2 \omega_{\dot{\alpha}}  d^2 \overline{\omega}^{\dot{\alpha}}  d\mu_{33}  d\overline{\mu}_{33}d^{N_1}\mu_{13}  d^{N_1}\overline{\mu}_{31}d^{N_2}\mu_{23}  d^{N_2}\overline{\mu}_{32} \nn \\
&&\phantom{\Lambda_{\mathrm{string}}^{3-N_1 -N_2} \int}\times ~\delta^{2}_{\mathrm{F}} \big(\ov\mu_{33} \omega_{ \dot\alpha} + \ov\omega_{\dot\alpha} \mu_{33}    \big)
~e^{ -S_1-S_2}~
\end{eqnarray}
where we have performed the integrals over the $\lambda^{\dot\alpha }$  variables in (\ref{S1}) and thereby implemented the fermionic ADHM constraints in terms of two $\delta$-functions. Let us also express $S_2$, from (\ref{S21}), in the following way
\begin{eqnarray}
\label{S22}
S_2 &=& \frac{1}{2} \overline{\omega}^{\dot{\alpha}} \{ \Phi \ov\Phi \}  \omega_{\dot{\alpha}}  
+\frac{i}{2}\overline{\mu}_{31}\ov\Phi_{13}\mu_{33} -\frac{i}{2}\overline{\mu}_{33}\ov\Phi_{31}\mu_{13}+\frac{i}{2}\overline{\mu}_{32}\ov\Phi_{23}\mu_{33}-\frac{i}{2}\overline{\mu}_{33}\ov\Phi_{32}\mu_{23} \nn \\
&&-\frac{i}{2}\overline{\mu}_{32}\Phi_{21}\mu_{13} + \frac{i}{2}\overline{\mu}_{31}\Phi_{12}\mu_{23} ~~
\end{eqnarray}
where $\{ \Phi \ov\Phi \} = \Phi_{31} \overline\Phi_{13} +\overline\Phi_{31} \Phi_{13} +\Phi_{32} \overline\Phi_{23} +\overline\Phi_{32} \Phi_{23} $.

\subsection{Fermionic Integration}

Due to the fermionic nature of the two $\delta$-functions brought down by the $\lambda$-integration, we can simply drop the ``$\delta_{\mathrm{F}}$'' and obtain the following two terms,
\begin{equation}
\label{adhmferm}
 \big(\ov\mu_{33} \omega_{ \dot{1}} + \ov\omega_{\dot{1}} \mu_{33}    \big)
\big(\ov\mu_{33} \omega_{ \dot{2}} + \ov\omega_{\dot{2}} \mu_{33}    \big)=
\ov\mu_{33}\Big( \omega_{\dot{1}}  \overline{\omega}_{\dot{2}} - \overline{\omega}_{\dot{1}}\omega_{\dot{2}}\Big) \mu_{33} =
\ov\mu_{33}\Big(\overline{\omega}^{\dot{1}} \omega_{\dot{1}}   + \overline{\omega}^{\dot{2}}\omega_{\dot{2}}\Big) \mu_{33}
\end{equation}
where we have raised indices using $\overline{\omega}_{\dot{\alpha}} =\epsilon_{\dot{\alpha}\dot{\beta}}\overline{\omega}^{\dot{\beta}}$, with  $\epsilon_{\dot{1}\dot{2}}=-\epsilon_{\dot{2}\dot{1}}=-1$. In (\ref{adhmferm}) we have also used the fact that the terms in which either $\mu_{33}$ or $\ov\mu_{33}$ appear twice vanish since these Grassmann variables  anticommute to zero. Since the terms in (\ref{adhmferm}) appear in front of the exponential in (\ref{int}), these terms soak up both $\mu_{33}$ and $\ov\mu_{33}$, implying that we are already done with the integration over these two variables. This means that, in order to get a non-vanishing result, we can only expand the terms in the exponent of (\ref{int}) containing $\mu_{33}$ and $\ov\mu_{33}$ to zeroth order. Thus, we forget about the last four couplings in the first line of (\ref{S22}) and instead study the last two couplings which include all the remaining charged fermionic moduli fields, $\overline{\mu}_{32}$, $\mu_{13}$, $\overline{\mu}_{31}$ and $\mu_{23}$. Since these remaining fields appear symmetrically we must expand both these terms to $N_{1}^{\ph{2}\mathrm{th}}$=$N_{2}^{\ph{2}\mathrm{th}}$ order to be able to soak up the remaining charged fermionic moduli fields. Hence, from the fermionic integration we get the constraint
\begin{equation}
\label{ }
N_1=N_2~.
\end{equation}  
The integration over the remaining fermions brings down determinants of $\Phi_{ 21}$ and $\Phi_{ 12}$ and we arrive at the following result,
\begin{eqnarray}
\label{int2}
W_{\mathrm{np}}^{\mathrm{s}}&=&  \Lambda_{\mathrm{string}}^{3-2N} \det[\Phi_{ 21}]\det[\Phi_{ 12}] \times\mathcal{I}~~~~~~~~~~~~~~\mathrm{for}~N_1=N_2=N~
\end{eqnarray}
where the remaining task is to evaluate the following bosonic integral,
\begin{eqnarray}
\label{intb}
\mathcal{I} =\int  d^3 D^c     d^2 \omega_{\dot{\alpha}}  d^2 \overline{\omega}^{\dot{\alpha}} \big( \overline{\omega}^{\dot{1}} \omega_{\dot{1}}   + \overline{\omega}^{\dot{2}}\omega_{\dot{2}}\big) 
~e^{i D^c ( \ov\omega^{ \dot\alpha}
(\tau^c)^{\dot\beta}_{\dot\alpha} \omega_{\dot\beta} ) - \frac{1}{2} \overline{\omega}^{\dot{\alpha}} \{ \Phi \ov\Phi \}  \omega_{\dot{\alpha}}    }~.
\end{eqnarray}
Note that dimensional analysis of (\ref{int2}) tells us that the bosonic integral $\mathcal{I}$ must be dimensionless. Hence, since $\mathcal{I}$ can only depend on the dimensionful quantity $\{ \Phi \ov\Phi \}$, we conclude that $\mathcal{I}$ must be a simple number, independent of $\{ \Phi \ov\Phi \}$. In the following section we will show that this number is non-zero.

\subsection{Bosonic Integration}

Since the charged bosonic moduli fields appear quadratically in the exponent of (\ref{intb}) we can simply insert the the  components of the three Pauli sigma matrices $\tau^c$ and arrive at the following expression for the bosonic integral,
\begin{eqnarray}
\label{int3}
\mathcal{I}=
\int   d^3 D   d^2 \omega  d^2 \overline{\omega} ~  \Big( -\frac{\partial}{\partial M_1} -\frac{\partial}{\partial M_4} \Big) ~  \mathrm{exp}\Bigg( -
\big[\begin{array}{cc} \overline{\omega}^{\dot{1}}& \overline{\omega}^{\dot{2}} \end{array}\big]
\left[\begin{array}{cc}M_1 & M_2 \\ M_3 & M_4 \end{array}\right]
\left[\begin{array}{c} \omega_{\dot{1}}\\ \omega_{\dot{2}} \end{array}\right]
 \Bigg) 
\end{eqnarray}
 where we have denoted $M_1 =- iD^3 + \frac{1}{2}\{ \Phi \overline\Phi \}$, $M_2 =-iD^1-D^2$, $M_3 =-iD^1+D^2$ and $M_4 =iD^3+\frac{1}{2}\{ \Phi \overline\Phi \}$.
Performing the Gaussian integrals over the charged bosonic moduli fields and taking the derivatives with respect to $M_1$ and $M_4$, we obtain
\begin{eqnarray}
\label{int4}
\int  d^3 D  \Big(- \frac{\partial}{\partial M_1} -\frac{\partial}{\partial M_4} \Big)  \frac{1}{M_1 M_4 - M_3 M_2} =\int   d^3 D  \frac{\{ \Phi \overline\Phi \} }{\big[ D^2 +\frac{1}{4}\{ \Phi \overline\Phi \}^2  \big]^{2}}  
 \end{eqnarray}
where we have inserted back the expressions for the $M$'s and denoted $D^2 = \sum_{c=1}^{3} (D^c)^2$. If we now change to spherical coordinates ($\int d^3 D =  4\pi \int dD ~D^2 $), rescale $\tilde{D}= \frac{2D}{\{ \Phi \overline\Phi \}} $ and use the fact that $\int_{0}^{\infty}   d \tilde{D} \frac{\tilde{D}^2}{[ \tilde{D}^2 +1 ]^2} =\frac{\pi}{4}$, we can conclude that the bosonic integral $\mathcal{I}$ from (\ref{intb}) only results in an irrelevant numerical factor which can be absorbed in the prefactor $\Lambda_{\mathrm{string}}$ in (\ref{int2}).

Thus, we have now shown that the final result from the complete moduli space integral was given in (\ref{int2}) and reads
\begin{eqnarray}
\label{int6}
W_{\mathrm{np}}^{\mathrm{s}}=  
\Lambda_{\mathrm{string}}^{3-2N} ~ \mathcal{B}\tilde{\mathcal{B}}
\end{eqnarray}
for $N_1=N_2=N$. In (\ref{int6}), we interpret the determinants from (\ref{int2}) as baryons, $\mathcal{B}=\det[\Phi_{21}]$ and $\tilde{\mathcal{B}}=\det[\Phi_{ 12}]$, and the superpotential term (\ref{int6}) as a stringy instanton generated baryonic mass term\footnote{Note that such a mass term can also be written as the determinant of the meson field,  $\det [\mathcal{M}] = \det[\Phi_{21}\Phi_{ 12}]$, and we will in fact show in the next section that the relation $\det [\mathcal{M}]= \mathcal{B}\tilde{\mathcal{B}}$ can be obtained as an equation of motion.}.
Note that we have generated a holomorphic superpotential term without using the D-term constraints for the matter fields, although we, of course, have to implement them in order to ensure supersymmetry.

Let us summarize our findings in slightly more general terms: We assumed that we had a background geometry with (at least) three non-trivial cycles, with two of them wrapped by space-filling D branes such that they, on their world volume, realized an $\mathcal{N}=1$ U($N$)$\times$U($N$) gauge theory with bi-fundamental matter. In that case, a non-perturbative superpotential term like (\ref{int6}) was generated by wrapping a single space-filling D brane together with an instanton ED brane on the third cycle. We believe that this result is quite general and should be applicable to many D brane systems in various backgrounds.

\section{Implications for the Gauge Dynamics}
\label{imp}

We have seen in the previous section that it is possible to generate a non-perturbative superpotential term for a U($N$)$\times$U($N$) gauge theory with an additional U(1) factor which has an instanton associated to it.  In this section we will discuss how we should expect the dynamics to change when we include the stringy instanton generated superpotential term (\ref{int6}).  

In an attempt to make contact  with the standard analysis of SQCD, 
let us for the moment consider the same configuration but in a background where we can take the limit where the volume of the second cycle, upon which $N$ D3 branes are wrapped, is large. We can then, using this limit, think of the case when the U($N$) group associated to the large cycle acts, together with the IR free U(1) factor from the third cycle, as a flavor group for the $N$ D3 branes wrapped on the first cycle that make up the U($N$) gauge group.  
This system is reminiscent of the $N_f =N_c +1$ (where $N_c =N$) SQCD case where we expect confinement but unbroken chiral symmetry  at the origin of the moduli space \cite{Seiberg:1994bz}. From a non-perturbative analysis of this specific SQCD theory we know that the classical moduli space, which in this case is the same as the quantum moduli space, is described by mesons $\mathcal{M}^{i}_{j}$ and baryons $\mathcal{B}_{i} $ and $\tilde{\mathcal{B}}^{j}$ where the flavor indices $i,j=1,..., N +1$, supplemented by certain constraints. These constraints can be implemented as equations of motion by the following non-perturbative superpotential term \cite{Seiberg:1994bz}, 
\begin{equation}
\label{nps}
W_{\mathrm{np}}= \Lambda^{-2N +1} \big(\mathcal{M}^{i}_{j} \mathcal{B}_{i} \tilde{\mathcal{B}}^{j} -\det[\mathcal{M}^{i}_{j}] \big)~.
\end{equation}

In order to make contact with our D brane configuration, which is a $\mathbb{Z}_2\times \mathbb{Z}_2$ orbifold projection of the $\mathcal{N}=4$ theory we must also remember to include the cubic tree level superpotential that explicitly breaks the flavor SU($N+1$)$\times$SU($N+1$) chiral symmetry of the SQCD theory, 
\begin{equation}
\label{treesup}
W_{\mathrm{tree}}= \Phi_{23}^{f}\Phi_{31 c}\Phi_{12 f}^c - \Phi_{21 c}^f \Phi_{13}^c \Phi_{32 f}
\end{equation} 
where the gauge indices $c=1,...,N$ and the flavor indices $f=1,..., N$. We will from now on only be interested in the part of the flavor group which has been broken down to SU$(N)\times$U(1) by (\ref{treesup}). Since our configuration is in the SQCD regime where $N_f =N_c +1$ we expect the presence of the non-perturbative superpotential in (\ref{nps}) although it will here describe fields with flavor indices decomposed into representations of SU$(N)\times$U(1). The mesons $\mathcal{M}^{i}_{j}$, which are in the $\mathbf{ad}+\mathbf{ 1}$ of SU($N +1$), are decomposed into the following representations of SU$(N)\times$U(1),
 \begin{eqnarray}
\label{mesde}
\mathcal{M}^{f}_{f'}= \Phi_{21 c}^{f}\Phi_{12 f'}^{c}~&:&~\mathrm{\mathbf{ad}}_0+\mathbf{ 1}_0 \nn \\
\mathcal{M}^{f}=\Phi_{21 c}^{f}\Phi_{13}^{c}~&:&~\mathrm{\mathbf{N} }_{-}\nn \\
\mathcal{M}_{f}=\Phi_{31 c}\Phi_{12 f}^{c}~&:&~\mathrm{ \mathbf{\ov{N}}}_{+}\nn \\
\mathcal{M}= \Phi_{31c}\Phi_{13}^{ c}~&:&~ \mathrm{\mathbf{1}}_{0}'~~.
\end{eqnarray}
 The baryons $\mathcal{B}_{i} $ and $\tilde{\mathcal{B}}^{j}$, in the $\ov{\mathbf{ N+1}}$ and $\mathbf{ N+1}$ of SU($N+1$), decompose according to
\begin{eqnarray}
\label{bade}
\mathcal{B}_f =\epsilon_{f f_1f_2 \cdots f_{N -1}}\epsilon^{c_1 \cdots c_{N}}\Phi_{31 c_1}
\Phi_{21 c_2}^{f_1}\cdots \Phi_{21 c_{N }}^{f_{N-1}}~&:&~\mathrm{ \mathbf{\ov{N}}}_+ \nn \\
\tilde{\mathcal{B}}^{f} =\epsilon^{f f_1f_2 \cdots f_{N -1}}\epsilon_{c_1 \cdots c_{N}}\Phi_{13}^{ c_1}\Phi_{12 f_1}^{c_2}\cdots \Phi_{12 f_{N-1}}^{c_{N}}~&:&~\mathrm{\mathbf{N} }_- \nn \\
\mathcal{B}=\epsilon_{f_1 \cdots f_N}\epsilon^{c_1 \cdots c_{N}}\Phi_{21 c_1}^{f_1} \cdots \Phi_{21 c_{N}}^{f_N}~&:&~ \mathrm{\mathbf{1}}_0 \nn \\
\tilde{\mathcal{B}}=\epsilon^{f_1 \cdots f_N}\epsilon_{c_1 \cdots c_{N}}\Phi_{12 f_1}^{c_1}\cdots \Phi_{12 f_{N}}^{c_{N}}~&:&~ \mathrm{\mathbf{1}}_0~~.
\end{eqnarray}

In order to see what effect the stringy instanton generated superpotential term from (\ref{int6}) might have on the SQCD theory described above, let us write out the superpotential in terms of the decomposed fields in (\ref{mesde}) and (\ref{bade}), and simply add the non-perturbative superpotential term  from (\ref{int6}), 
\begin{eqnarray}
\label{bbt}
W_{\mathrm{tree}}+W_{\mathrm{np}}+W_{\mathrm{np}}^{\mathrm{s}}&=&
\Big( \Lambda^{-2N +1} \mathcal{M}  +\Lambda_{\mathrm{string}}^{-2N +3}\Big) \mathcal{B} \tilde{\mathcal{B}} -\Lambda^{-2N +1} \mathcal{M} \det [\mathcal{M}^{f}_{f'}]+\cdots ~.
\end{eqnarray}
The dots in (\ref{bbt}) refer to terms which are not important for our discussion, for example, those that include the fields $\mathcal{M}^{f}$ and $\mathcal{M}_{f}$ which are both set to zero by the equations of motion for the fields $\Phi_{23}^{f}$ and $\Phi_{32 f}$, see (\ref{treesup}). Note that the stringy term affects only the term from (\ref{nps}) which is of the same form and the effect can be seen as a shift in the flavor singlet field $\mathcal{M}$, without removing the moduli space singularities. Further note that if we interpret the stringy superpotential from (\ref{int6}) as $\det [\mathcal{M}^{f}_{f'}]$, instead of $\mathcal{B} \tilde{\mathcal{B}}$, we get a similar shifting effect for $\mathcal{M}$,
\begin{eqnarray}
\label{mm}
W_{\mathrm{tree}}+W_{\mathrm{np}}+W_{\mathrm{np}}^{\mathrm{s}}&=& 
 \Big( -\Lambda^{-2N +1}\mathcal{M}  +\Lambda_{\mathrm{string}}^{-2N +3} \Big) \det [\mathcal{M}^{f}_{f'}] +\Lambda^{-2N +1} \mathcal{M}~ \mathcal{B} \tilde{\mathcal{B}} \cdots~.
\end{eqnarray}
Regardless of how we interpret the stringy superpotential in (\ref{int6}), the equation of motion for $\mathcal{M}$ gives the constraint, $\det [\mathcal{M}^{f}_{f'}] =\mathcal{B} \tilde{\mathcal{B}} $. 

From (\ref{bbt}) we see that the stringy instanton would break R-symmetry unless we can assign R charge zero to the field $\mathcal{M}$. And as expected in this non-chiral gauge theory, there is a non-anomalous R-charge assignment such that the fields $ \Phi_{31c}$ and $\Phi_{13}^{ c}$ have R-charge zero, see Table 2. Thus, by including the stringy instanton, we fix the the non-anomalous U(1$)_{\mathrm{R}}$ current uniquely. Moreover, there is a non-anomalous axial U(1$)_{\mathrm{A}}$ symmetry, under which the charge of $\mathcal{B} \tilde{\mathcal{B}}$ is compensated by the opposite charge of $\mathcal{M}$, but which is here non-perturbatively broken by the stringy instanton.
\begin{table} [ht]
\begin{center}
\begin{tabular}{c||c|c|c|}
&$U(1)_{\mathrm{R}}$  \\ 
\hline   \hline  $\Phi_{12}$  & $\frac{1}{N}$ \\  
\hline $\Phi_{21}$ & $\frac{1}{N}$ \\ 
 \hline $\Phi_{13}$ &0 \\
 \hline $\Phi_{31}$ &0 \\
  \hline $\Phi_{23}$ & 2-$\frac{1}{N}$ \\
   \hline $\Phi_{32}$ & 2-$\frac{1}{N}$\\
  \end{tabular}
\caption{\small The non-anomalous R-charge assignment.}
\end{center}
\end{table}

In conclusion, we have in this paper shown that it is possible to generate an interesting non-perturbative superpotential term (\ref{int6}) by a stringy instanton in a simple $\mathbb{Z}_2\times \mathbb{Z}_2$ orbifold background without introducing orientifold planes. In more generic Calabi-Yau backgrounds, it has been shown that similar terms as the one we have found here will have dramatic effects on the gauge dynamics and, in some cases, give rise to dynamical supersymmetry breaking \cite{Argurio:2007qk,Aharony:2007db,Buican:2007is,Aharony:2007pr}. We regard this computation as an example of how such an effect could arise in more realistic theories.


\section*{Acknowledgements}

It is a pleasure to thank Riccardo Argurio, Gabriele Ferretti, Alberto Lerda and Daniel Persson for fruitful discussions and for reading the manuscript. The author would also like to thank Matteo Bertolini, Jose F. Morales, Bengt E.W. Nilsson and Niclas Wyllard for interesting conversations.


\begin{thebibliography}{99}

\bibitem{Witten:1995gx} E.~Witten, 
Nucl.\ Phys.\  B {\bf 460} (1996) 541
  [arXiv:hep-th/9511030].  

\bibitem{Douglas:1995bn} M.~R.~Douglas, 
  [arXiv:hep-th/9512077].  

\bibitem{Witten:1996bn} E.~Witten, 
Nucl.\ Phys.\  B {\bf 474}, 343 (1996)
  [arXiv:hep-th/9604030].  

\bibitem{Ganor:1996pe} O.~J.~Ganor,
Nucl.\ Phys.\  B {\bf 499}, 55 (1997)
  [arXiv:hep-th/9612077].  

\bibitem{Green:2000ke} M.~B.~Green and M.~Gutperle,
JHEP {\bf 0002} (2000) 014 [arXiv:hep-th/0002011].  

\bibitem{Billo:2002hm} M.~Billo, M.~Frau, I.~Pesando, F.~Fucito,
  A.~Lerda and A.~Liccardo, 
JHEP {\bf 0302} (2003) 045 [arXiv:hep-th/0211250].

\bibitem{Akerblom:2006hx} N.~Akerblom, R.~Blumenhagen, D.~Lust,
  E.~Plauschinn and M.~Schmidt-Sommerfeld,
[arXiv:hep-th/0612132].

\bibitem{Dorey:2002ik} N.~Dorey, T.~J.~Hollowood, V.~V.~Khoze and
  M.~P.~Mattis, 
Phys.\ Rept.\ {\bf 371} (2002) 231 [arXiv:hep-th/0206063].

\bibitem{Beasley:2004ys}
  C.~Beasley and E.~Witten,
  JHEP {\bf 0501} (2005) 056
  [arXiv:hep-th/0409149].
  
\bibitem{Blumenhagen:2006xt} R.~Blumenhagen, M.~Cvetic and T.~Weigand,
  [arXiv:hep-th/0609191].

\bibitem{Ibanez:2006da} L.~E.~Ibanez and A.~M.~Uranga, 
  [arXiv:hep-th/0609213].  

\bibitem{Florea:2006si} B.~Florea, S.~Kachru, J.~McGreevy and
  N.~Saulina, 
  [arXiv:hep-th/0610003].  

\bibitem{Abel:2006yk}
  S.~A.~Abel and M.~D.~Goodsell,
  [arXiv:hep-th/0612110].

\bibitem{Bianchi:2007fx} M.~Bianchi and E.~Kiritsis,
  [arXiv:hep-th/0702015].  

\bibitem{Cvetic:2007ku} M.~Cvetic, R.~Richter and T.~Weigand,
  [arXiv:hep-th/0703028].  
  
   \bibitem{Argurio:2007qk}
  R.~Argurio, M.~Bertolini, S.~Franco and S.~Kachru,
  JHEP {\bf 0706}, 017 (2007)
  [arXiv:hep-th/0703236].
  
 \bibitem{Argurio:2007vqa}
  R.~Argurio, M.~Bertolini, G.~Ferretti, A.~Lerda and C.~Petersson,
  JHEP {\bf 0706}, 067 (2007)
  [arXiv:0704.0262 [hep-th]].


\bibitem{Bianchi:2007wy}
  M.~Bianchi, F.~Fucito and J.~F.~Morales,
  JHEP {\bf 0707}, 038 (2007)
  [arXiv:0704.0784 [hep-th]].
  
\bibitem{Ibanez:2007rs}
  L.~E.~Ibanez, A.~N.~Schellekens and A.~M.~Uranga,
  JHEP {\bf 0706} (2007) 011
  [arXiv:0704.1079 [hep-th]].
  
\bibitem{Blumenhagen:2007sm}
  R.~Blumenhagen, S.~Moster and E.~Plauschinn,
  [arXiv:0711.3389 [hep-th]].
  
\bibitem{Antusch:2007jd}
  S.~Antusch, L.~E.~Ibanez and T.~Macri,
  JHEP {\bf 0709}, 087 (2007)
  [arXiv:0706.2132 [hep-ph]].
    
\bibitem{Franco:2007ii}
  S.~Franco, A.~Hanany, D.~Krefl, J.~Park, A.~M.~Uranga and D.~Vegh,
  JHEP {\bf 0709} (2007) 075
  [arXiv:0707.0298 [hep-th]].
  
\bibitem{Ibanez:2007tu}
  L.~E.~Ibanez and A.~M.~Uranga,
  [arXiv:0711.1316 [hep-th]].
    
    
  \bibitem{Blumenhagen:2007bn}
  R.~Blumenhagen, M.~Cvetic, R.~Richter and T.~Weigand,
  JHEP {\bf 0710}, 098 (2007)
  [arXiv:0708.0403 [hep-th]].

\bibitem{Martucci:2005rb}
  L.~Martucci, J.~Rosseel, D.~Van den Bleeken and A.~Van Proeyen,
  ``Dirac actions for D-branes on backgrounds with fluxes,''
  Class.\ Quant.\ Grav.\  {\bf 22}, 2745 (2005)
  [arXiv:hep-th/0504041].

\bibitem{Bergshoeff:2005yp}
  E.~Bergshoeff, R.~Kallosh, A.~K.~Kashani-Poor, D.~Sorokin and A.~Tomasiello,
  ``An index for the Dirac operator on D3 branes with background fluxes,''
  JHEP {\bf 0510}, 102 (2005)
  [arXiv:hep-th/0507069].

\bibitem{Haack:2006cy}
  M.~Haack, D.~Krefl, D.~Lust, A.~Van Proeyen and M.~Zagermann,
  JHEP {\bf 0701} (2007) 078
  [arXiv:hep-th/0609211].

\bibitem{Franco:2005zu} S.~Franco, A.~Hanany, F.~Saad and
  A.~M.~Uranga, 
JHEP {\bf 0601} (2006) 011 [arXiv:hep-th/0505040].





\bibitem{Bertolini:2005di}
  M.~Bertolini, F.~Bigazzi and A.~L.~Cotrone,
  ``Supersymmetry breaking at the end of a cascade of Seiberg dualities,''
  Phys.\ Rev.\  D {\bf 72} (2005) 061902
  [arXiv:hep-th/0505055].
  
\bibitem{Berenstein:2005xa}
  D.~Berenstein, C.~P.~Herzog, P.~Ouyang and S.~Pinansky,
  JHEP {\bf 0509} (2005) 084
  [arXiv:hep-th/0505029].

 
  
 
  
    
\bibitem{Atiyah:1978ri} M.~F.~Atiyah, N.~J.~Hitchin, V.~G.~Drinfeld
  and Yu.~I.~Manin, 
Phys.\ Lett.\  A  {\bf 65}, 185 (1978).  

\bibitem{Aganagic:2007py}
  M.~Aganagic, C.~Beem and S.~Kachru,
 [arXiv:0709.4277 [hep-th]].
  
\bibitem{GarciaEtxebarria:2007zv}
  I.~Garcia-Etxebarria and A.~M.~Uranga,
  [arXiv:0711.1430 [hep-th]].


  
  
  \bibitem{Aharony:2007db}
  O.~Aharony, S.~Kachru and E.~Silverstein,
 [arXiv:0708.0493 [hep-th]].
  
\bibitem{Buican:2007is}
  M.~Buican, D.~Malyshev and H.~Verlinde,
  [arXiv:0710.5519 [hep-th]].
    
\bibitem{Aharony:2007pr}
  O.~Aharony and S.~Kachru,
  JHEP {\bf 0709}, 060 (2007)
  [arXiv:0707.3126 [hep-th]].
  
  
  \bibitem{Bianchi:2007ft} M.~Bianchi, S.~Kovacs and G.~Rossi,
[arXiv:hep-th/0703142].
  
   
   \bibitem{Witten:1995im} E.~Witten,
Nucl.\ Phys.\  B {\bf 460} (1996) 335
  [arXiv:hep-th/9510135].  
  
  
  \bibitem{Brink:1976bc}
  L.~Brink, J.~H.~Schwarz and J.~Scherk,
  Nucl.\ Phys.\  B {\bf 121}, 77 (1977).
  
  \bibitem{Douglas:1996sw} M.~R.~Douglas and G.~W.~Moore,
[arXiv:hep-th/9603167].
  
\bibitem{Morrison:1998cs}
  D.~R.~Morrison and M.~R.~Plesser,
  Adv.\ Theor.\ Math.\ Phys.\  {\bf 3} (1999) 1
  [arXiv:hep-th/9810201].
    
  \bibitem{Bertolini:2001gg} M.~Bertolini, P.~Di Vecchia, G.~Ferretti
  and R.~Marotta, 
  Nucl.\ Phys.\ B {\bf 630} (2002) 222 [arXiv:hep-th/0112187].

\bibitem{Bertolini:2003iv}
  M.~Bertolini,
  Int.\ J.\ Mod.\ Phys.\  A {\bf 18} (2003) 5647
  [arXiv:hep-th/0303160].
  
   \bibitem{Taylor:1982bp} T.~R.~Taylor, G.~Veneziano and
  S.~Yankielowicz, 
  Nucl.\ Phys.\  B {\bf 218} (1983)
  493.  

\bibitem{Affleck:1983mk} I.~Affleck, M.~Dine and N.~Seiberg,
  Phys.\  B {\bf 241} (1984) 493.  
  
  
   \bibitem{Akerblom:2007uc}
  N.~Akerblom, R.~Blumenhagen, D.~Lust and M.~Schmidt-Sommerfeld,
  JHEP {\bf 0708}, 044 (2007)
  [arXiv:0705.2366 [hep-th]].
  
  
  
 \bibitem{Billo:2007py}
  M.~Billo, M.~Frau, I.~Pesando, P.~Di Vecchia, A.~Lerda and R.~Marotta,
 [arXiv:0709.0245 [hep-th]].
  
   \bibitem{Billo:2007sw}
  M.~Billo, M.~Frau, I.~Pesando, P.~Di Vecchia, A.~Lerda and R.~Marotta,
  JHEP {\bf 0710}, 091 (2007)
  [arXiv:0708.3806 [hep-th]].
  
    
   
\bibitem{Lust:2003ky}
  D.~Lust and S.~Stieberger,
  Fortsch.\ Phys.\  {\bf 55}, 427 (2007)
  [arXiv:hep-th/0302221].
    
\bibitem{Akerblom:2007np}
  N.~Akerblom, R.~Blumenhagen, D.~Lust and M.~Schmidt-Sommerfeld,
  Phys.\ Lett.\  B {\bf 652}, 53 (2007)
  [arXiv:0705.2150 [hep-th]].
  
    
 
\bibitem{Camara:2007dy}
  P.~G.~Camara, E.~Dudas, T.~Maillard and G.~Pradisi,
  [arXiv:0710.3080 [hep-th]].
  
  \bibitem{Polchinski:1994fq}
  J.~Polchinski,
  Phys.\ Rev.\  D {\bf 50}, 6041 (1994)
  [arXiv:hep-th/9407031].
    

    
    
\bibitem{Martin:1999aq}
  C.~P.~Martin and D.~Sanchez-Ruiz,
  Phys.\ Rev.\ Lett.\  {\bf 83} (1999) 476
  [arXiv:hep-th/9903077].
  
\bibitem{Hayakawa:1999zf}
  M.~Hayakawa,
  [arXiv:hep-th/9912167].
 
  
\bibitem{Khoze:2000sy}
  V.~V.~Khoze and G.~Travaglini,
  JHEP {\bf 0101}, 026 (2001)
  [arXiv:hep-th/0011218].
    
\bibitem{Nekrasov:1998ss}
  N.~Nekrasov and A.~S.~Schwarz,
  Commun.\ Math.\ Phys.\  {\bf 198} (1998) 689
  [arXiv:hep-th/9802068].
    
    
  \bibitem{Seiberg:1994bz}
  N.~Seiberg,
  Phys.\ Rev.\  D {\bf 49}, 6857 (1994)
  [arXiv:hep-th/9402044].
  
  
  
  
  
\end{thebibliography}
\end{document}